\documentclass[]{article}
\usepackage{autoref}
\begin{document}
\author{Галлямов\,\, Марат\,\, Олегович}
\title{СКАНИРУЮЩАЯ ЗОНДОВАЯ МИКРОСКОПИЯ НУКЛЕИНОВЫХ КИСЛОТ
И ТОНКИХ ОРГАНИЧЕСКИХ ПЛЕНОК}
\date{June 1999}

\maketitle

\section*{ОБЩАЯ ХАРАКТЕРИСТИКА РАБОТЫ}

\paragraph{Актуальность темы.} 

В последние годы методы сканирующей зондовой микроскопии позволили достичь уникальных научных результатов в различных областях физики, химии и биологии. Новые экспериментальные возможности данного направления~\emdash неразрушающий характер исследований, высокое пространственное разрешение и возможность проведения экспериментов в жидких средах~\emdash делают особенно перспективным применение СЗМ\footnote{СЗМ~\emdash сканирующая зондовая микроскопия} (в частности, атомно-силовой микроскопии) для изучения структуры и свойств \emph {биологических и органических} материалов. 
В то же время СЗМ-исследование этих объектов остается более сложной задачей в сравнении с аналогичными исследованиями поверхностей твердых (кристаллических) тел. Действительно, прошло более десяти лет с момента возникновения зондовой микроскопии (в 1981\,г.), прежде чем в 1992\,г. была убедительно продемонстрирована адекватность этого метода для исследований биополимеров на примере АСМ\footnote{АСМ~\emdash атомно-силовая микроскопия}-визуализации молекулы ДНК\footnote{ДНК~\emdash дезоксирибонуклеиновая кислота}. За эти годы возникло понимание, что \emph{определяющей задачей} успешности подобных исследований, требующей экспериментального решения в каждом конкретном случае, является \emph{иммобилизация} биологических и органических структур на поверхностях твердых подложек в таком состоянии, чтобы было возможным исследовать их структурные особенности.

Весьма важным для адекватного применения СЗМ в широкомасштабных научных исследованиях является отслеживание и систематизация возможных механизмов возникновения \emph{артефактов}, т.е. аппаратных эффектов, приводящих к наблюдению ложных или искаженных свойств исследуемого объекта. 

\paragraph {Цель и задачи работы.}

Целью диссертации являлась разработка методов СЗМ-исследования нуклеиновых кислот и их комплексов, что и определило основные задачи:
\begin{itemize}
\item разработка алгоритмов определения реальной геометрии объекта из анализа экспериментально измеренных параметров его АСМ-изображения (учет артефактов);
\item определение адекватных методик иммобилизации молекул нуклеиновых кислот и их комплексов с белками и поверхностно-активными веществами в различных экспериментальных условиях;
\item отработка методов контролируемой модификации свойств подложки нанесением тонких органических пленок; исследование влияния природы подложки и процедуры формирования пленок на структуру покрытия.
\end{itemize}

\paragraph{Материалы и методы.} АСМ-исследования проводили в режимах постоянного или прерывистого\footnote{TappingMode$^{\mbox{\tiny TM}}$} контактов на приборе \lk Nanoscope-IIIa\pk\ (Digital Instruments, США) с использованием коммерческих Si или Si$_3$N$_4$ \emph{кантилеверов} (Nanoprobe, Digital Instruments, США). Эксперименты в жидких средах проводили с использованием жидкостной ячейки (Digital Instruments, США). В качестве подложек использовали слюду (мусковит) или высокоориентированный пиролитический графит (пирографит): как поверхности свежего скола, так и модифицированные обработкой катионными ПАВ\footnote{ПАВ~\emdash поверхностно-активные вещества}, катионами металлов, аминопропилтриэтоксисиланом или нанесением тонкопленочного органического покрытия. Обработку и построение АСМ-изображений проводили при помощи программного обеспечения \lk Nanoscope-IIIa\pk\ (Digital Instruments, США) и \lk Фемтоскан-001\pk\ (Центр перспективных технологий, Россия). 

При приготовлении всех препаратов использовали бидистиллированную деионизованную воду, контроль чистоты воды проводили методом АСМ. При исследованиях нуклеиновых кислот и их комплексов на воздухе каплю препарата наносили на поверхность подложки, экспонировали, промывали и высушивали в естественной атмосфере. При исследовании в жидких средах препарат вводили в жидкостную ячейку прибора и исследовали структуры, адсорбировавшиеся из раствора на поверхность подложки. 

При приготовлении пленок Ленгмюра-Блоджетт использовали автоматизированную ЛБ\footnote{ЛБ~\emdash Ленгмюра-Блоджетт} ванну, при этом формировали тонкопленочные покрытия как вертикальным методом ЛБ, так и методом горизонтального осаждения; давление выделения $\pi$ поддерживали на уровне 20--40\,мН/м. При АСМ-анализе параметров молекулярной упаковки тонких пленок специальное внимание уделили исключению искажающего влияния температурного дрейфа.

\paragraph{Научная новизна и практическая ценность работы.} 

Разработаны количественные методики описания двух искажающих эффектов АСМ: \emph{уширения} и \emph{занижения профиля} изображений отдельных микрообъектов, адсорбированных на поверхность твердой подложки. Предложен простой в реализации алгоритм численного решения, позволяющий восстановить их истинные размеры. Проанализирован механизм достижения \lk атомного\pk\ разрешения в АСМ в свете результатов теории контактных деформаций. На основе соотношений Герца этой теории разработана методика определения модуля упругости микрообъектов, иммобилизованных на подложке, с использованием АСМ. 

Апробирована методика приготовления образцов для АСМ-исследования молекул однонитевой РНК\footnote{РНК~\emdash рибонуклеиновая кислота} и их взаимодействия с белками. Применение данной методики позволило впервые методом АСМ визуализировать стадии процесса высвобождения молекулы РНК из белковой оболочки частиц ВТМ\footnote{ВТМ~\emdash вирус табачной мозаики}. При этом, в отличие от исследований электронной микроскопии, методика \emph{не включает} процесс какого-либо дополнительного контрастирования молекул РНК (комплексоообразования с белковой пленкой, запыления и т.п.). Проведение подобных АСМ-исследований в условиях близких нативным может дать вклад в понимание механизмов процесса размножения вирусов.

Динамика процессов компактизации/декомпактизации молекул высокомолекулярной ДНК~Т4 впервые исследована методом АСМ непосредственно в водно-спиртовой смеси (при контролируемом изменении концентрации изопропанола). В сравнении с результатами флуоресцентной микроскопии удалось достичь существенно более высокого пространственного разрешения, что позволило визуализировать тороидальные образования, сформированные отдельными витками частично компактизованных молекул. По результатам наблюдений и анализа геометрии компактной глобулы предложена модель компактизации ДНК. Разработанный подход может позволить исследовать системы, моделирующие транспорт генетического материала внутрь живых клеток. 

В экспериментах по АСМ-визуализации структуры пленок ЛБ показано преимущество метода горизонтального осаждения монослоев на подложку. С помощью АСМ анализировали влияние подложки на молекулярную упаковку тонких пленок. Проведение подобных исследований может позволить прояснить основные механизмы, определяющие формирование структуры пленки.

По разработкам диссертации поставлена задача спец. практикума кафедры физики полимеров и кристаллов физического факультета МГУ \lk Сканирующая зондовая микроскопия нуклеиновых кислот\pk. 

\paragraph{Апробация работы.} 

Основные результаты диссертации были доложены:\protect\\
\emph{на международных конференциях}: {1-й} и {2-й} по химии высокоорганизованных веществ и научным основам нанотехнологии (Санкт-Петербург, июнь 1996 и 1998\,гг.), 4-й и 5-й (\lk NANO-4\pk\ и \lk -5\pk) по наноразмерной науке и технологии (Пекин, Китай, сентябрь 1996 и Бирмингем, Великобритания, август 1998), \lk Фундаментальные проблемы науки о полимерах (к 90-летию академика В.\,А.\,Каргина)\pk\ (Москва, Россия, январь 1997), \lk Nanomeeting-97\pk\ и \lk -99\pk\ (Минск, Беларусь, май 1997 и 1999), 9-й по сканирующей туннельной микроскопии, спектроскопии и близким технологиям (Гамбург, Германия, июль 1997), пользователей NanoScope 1997 года (Санта-Барбара, США, август 1997);\qquad
\emph{на международных симпозиумах}: 7-м ежегодном \lk Фотоиндуцированный зарядовый перенос: реакции в живой материи\pk\ (Рочестер, США, июнь 1996), по коллоидной науке и науке о полимерах \lk Формирование и динамика самоорганизованных структур в растворах полимеров и поверхностно-активных веществ~\emdash последние достижения\pk\ (Нагойа, Япония, октябрь 1996), 43-м национальном американского вакуумного общ-ва \lk Программа наноразмерной науки и технологии\pk\ (Филадельфия, США, октябрь 1996);\qquad
на XVI Менделеевском съезде по общей и прикладной химии (Санкт-Петербург, Россия, май 1998), на всероссийских рабочих совещаниях \lk Зондовая микроскопия-97\pk, \lk -98\pk\ и \lk -99\pk\, (Нижний Новгород, Россия, март 1997, 1998 и 1999\,гг.); на 2-м и 3-м Белорусских семинарах по СЗМ (Минск, Беларусь, май 1997 и Гродно, Беларусь, октябрь 1998), на Школе по химии и физике полимеров (Тверь, Россия, декабрь 1998). 

По результатам работы опубликовано 6 статей и 20 тезисов и рефератов докладов, сделанных на конференциях. В список литературы, представленный в автореферате, вошли статьи по теме диссертации, а также рефераты докладов, материалы которых не нашли полного отражения в опубликованных статьях.

\paragraph{Личный вклад автора.} Все экспериментальные измерения зондовой микроскопии, разработка и применение теоретических моделей для интерпретации и анализа экспериментальных данных выполнены автором лично. Автор принимал участие в совместной работе с научными группами Химического факультета МГУ по приготовлению образцов для исследований.

\paragraph{Структура диссертации.} Диссертация состоит из введения, общего обзора литературы, теоретической и экспериментальной части, выводов, библиографии (132~наименования) и приложения. Теоретическая часть состоит из одной, а экспериментальная из трех глав; каждая глава содержит краткий литературный обзор по конкретной теме исследования. Работа изложена на 227~стр. (с приложением), содержит 54~рисунка и 9~таблиц. 

\bigskip

\section*{ОСНОВНОЕ СОДЕРЖАНИЕ РАБОТЫ}
\section{Основные принципы СЗМ}

В первой главе дан аналитический обзор основных методов и способов измерений зондовой микроскопии, ее возможностей и ограничений. Проанализировано место СЗМ в ряду других физических методов исследования \emph{поверхности}, рассмотрены общие подходы к препарированию образцов нуклеиновых кислот для СЗМ-исследований.

\section{Анализ искажающих эффектов АСМ}
\label{chap2}

Результатом работы АСМ является АСМ-изображение: профиль перемещения зонда в процессе \emph{сканирования} при фиксации системой обратной связи постоянного силового взаимодействия зонда и образца. АСМ-изображения могут неадекватно отображать реальную топографию исследуемой поверхности (говорят о наблюдении \emph{артефактов}). В работе предложены количественные описания двух основных артефактов АСМ, проявляющихся при исследовании объектов, адсорбированных на поверхность твердой подложки~\emdash \emph{эффектов уширения профиля} и \emph{ занижения высот} их АСМ-изображений.

\section* {Контактные деформации зонда и образца}
\label{con_def}

При сканировании зонд воздействует на образец с силой в \emph{единицы--десятки наноньютонов}, что, в связи с малым радиусом кривизны кончика зонда (около $10\,\mbox{нм}$), приводит к значительному \emph{контактному} давлению\footnote{по нашим оценкам: до 1~ГПа и более}, которое должно вызывать контактные деформации исследуемого объекта и приводить к занижению высоты его АСМ-изображения.
Величина деформации контактирующих тел определяется известными соотношениями \emph{контактной задачи Герца}\footnote{см., например, Л.\,Д.\,Ландау, Е.\,М.\,Лифшиц. \emph{Теория упругости.} \emdash М.:\,Наука, 1987}, которые показывают, что формой области контакта двух тел, сдавливаемых силой $F$, является эллипс с полуосями $a$ и $b$, причем величина сближения деформируемых тел $h$ может быть выражена через $F$, $a$, $b$, упругий параметр области контакта $D$\footnote{
$ D = {3}/{4}\left [ ({1-\sigma ^2})/{E}
+({1-\sigma '^2})/{E'}\right ],$
здесь $E$, $E'$, $\sigma$ и $\sigma'$~\emdash модули Юнга и Пуассона материалов зонда и образца
}, а также главные значения суммарного тензора кривизны контактирующих поверхностей~\emdash $A$ и $B$, которые определяются геометрией контакта\footnote{для контакта сферы радиуса $R_{зонд}$ и боковой поверхности цилиндра радиуса $R_{обр}$ они выражаются:
$
A={1}/{2}\,({1}/{R_{зонд}}+{1}/{R_{обр}}),
$
$
B={1}/{2R_{зонд}}
$
}. 

\subsubsection*{Контакт сферического зонда и цилиндрического образца.}

Модель цилиндрического образца находит применение при анализе деформаций микрочастиц цилиндрической формы (вирусных частиц, линейных макромолекул и пр.). Однако в этом случае\footnote{в отличие от известного случая \emph{сферической} геометрии контактирующих тел} соотношения Герца, включающие систему нелинейных уравнений с неявными зависимостями от искомых параметров, напрямую не упрощаются. Поэтому реализовали общее численное решение, сведя исходные соотношения к независимым уравнениям относительно \emph{одного} неизвестного. 

Для двух частных случаев удалось получить \emph{аналитические} формулы, выражающие искомые параметры явно: 

а)\quad Если $R_{обр}<R_{зонд}$, то главные значения суммарного тензора кривизны различаются: $A>B$. Тогда $a<b$ и, если, $a^2\ll b^2$, то можно показать, что: 
\begin{equation}
\label{t_hhh1}
h\simeq \left(\frac{4}{\pi^2 C}\right)^{1/3}(С+1)\times (FD)^{2/3}\times B^{1/3},
\end{equation}
где безразмерный параметр $C$ (зависящий от $a/b$) для многих задач лежит в диапазоне от 1 до 3, и с достаточной точностью можно воспользоваться оценкой.

б)\quad Если $R_{обр} \gg R_{зонд}$, то $A\sim B$, тогда $a\sim b $ и можно показать:
\begin{equation}
\label{th_h_01}
h\simeq (FD)^{2/3}\times \left(\frac{1}{4A}+\frac{1}{4B}\right)^{-1/3},
\end{equation}
что имеет структуру, сходную с~(\ref{t_hhh1}). 
В обоих случаях оказалось возможным получить формулы и для параметров эллиптической области контакта $a$ и $b$, что позволяет определить контактное давление.

\begin{figure}[!h]
\begin{center}
\includegraphics[width=0.6\textwidth]{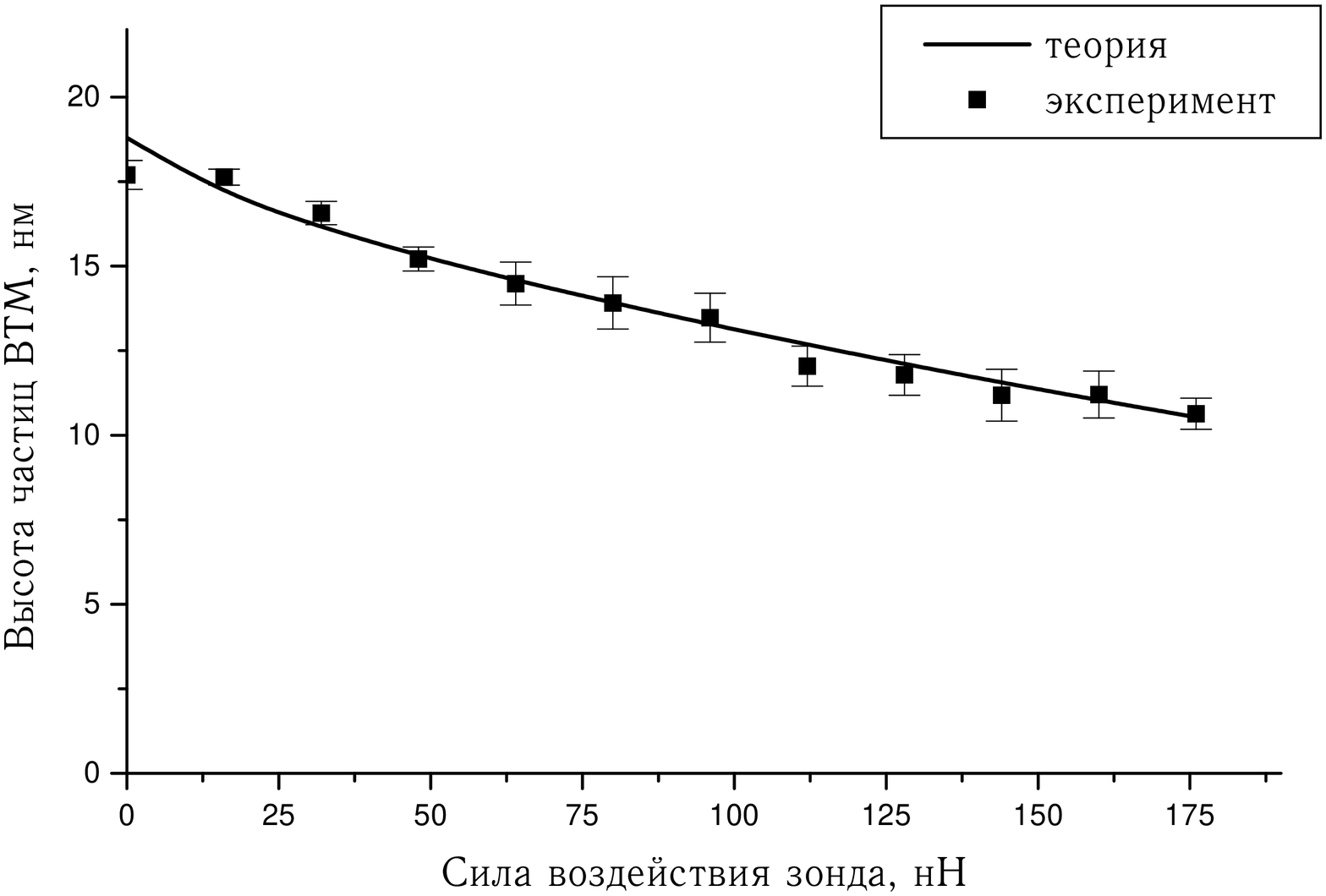}
\parbox[b]{0.39\textwidth}{
\caption{Экспериментальная и рассчитанная по соотношениям Герца зависимости высоты профиля АСМ-изо\-бра\-же\-ний частиц вируса табачной мозаики от величины нагружающей силы при сканировании. }
\label{def_vtm1}
}
\vskip -0.5cm
\end{center}
\end{figure}

Результаты расчетов предсказывают тем большую величину относительных деформаций объектов, чем меньше их радиус кривизны, что подтверждается экспериментом. С целью экспериментальной проверки закона $h\sim F^{2/3}$ (см.~(\ref{t_hhh1}), (\ref{th_h_01})) мы провели исследования зависимости деформации частиц ВТМ\footnote{частицы вируса табачной мозаики имеют цилиндрическую форму} от силы воздействия зонда и обнаружили хорошее совпадение эксперимента с результатами, рассчитанными по разработанному подходу, см. рис.\,\ref{def_vtm1}. Анализ экспериментальной зависимости позволил определить модуль упругости отдельной вирусной частицы ($E\sim 3\div 4\,\mbox{ГПа}$).

\subsubsection*{Обоснование возможности достижения атомного разрешения в АСМ.}

\label{m_r_desc}

Результаты применения теории контактных деформаций для контакта зонда с радиусом кривизны кончика~$R$ и плоского образца показывают, что при типичных условиях АСМ-эксперимента на воздухе радиус области контакта зонда и образца\footnote{$a=(FDR)^{1/3}$} оказывается $\sim 1\,\mbox{нм}$ и \emph{превышает} межатомное расстояние. Это означает, что при сканировании в каждый момент времени с зондом взаимодействуют \emph{несколько} атомов образца. Возникает вопрос о механизме АСМ-визуализации атомной структуры. Действительно, контакт с \emph{единственным} атомом ($a\sim$ межатомного расстояния) может быть осуществлен лишь при минимизации силы воздействия зонда до величины $10\div100$\,пН, что требует специальных экспериментальных условий\footnote{например, проведения исследований в жидкости}.

Известны наблюдаемые в ряде случаев особенности АСМ-визуализации атомной структуры поверхности: инверсия контраста, визуализация \lk ложного атома\pk\ на месте точечного дефекта. Ранее эти наблюдения объяснялись с позиции наличия на \emph{кончике иглы} нескольких атомов, дающих одинаковый вклад во взаимодействие с образцом. Мы предлагаем подход, связывающий эти особенности с параметрами эксперимента, определяющими радиус контактной площадки~$a$.

В силу \emph{неоднородного распределения давления} в области контакта вклад каждого атома образца в силовое взаимодействие с зондом будет определяться его положением относительно центра области контакта. Формула контактной задачи Герца, описывающая распределение контактного давления, позволяет ввести аппаратную функцию АСМ\footnote{ее применимость ограничена случаем плоского образца и сферического кончика зонда} вида:
\begin{equation}
\label{appar2}
\begin{array}{c}
\displaystyle
A(x-x',y-y')\simeq \\
\simeq
\left\{
\begin{array}{ll}
\displaystyle
\sqrt{1-\frac{(x-x')^2+(y-y')^2}{a^2}} & \mbox{если~} 
(x-x')^2+(y-y')^2 \leq a^2\\
\displaystyle
0 & \mbox{если~} (x-x')^2+(y-y')^2 > a^2,
\end{array}
\right. \\
\end{array}
\end{equation}
где $a$\,\emdash радиус области контакта. Введенная аппаратная функция связывает АСМ-изображение $f(x,y)$ с реальной геометрией атомной решетки поверхности исследования $\varphi(x,y)$:
\begin{equation}
\label{appar3}
f(x,y)=\int \!\! \int A(x-x',y-y')\varphi(x',y')\, dx'dy'
\end{equation} 

\begin{figure}
\begin{center}
\parbox[c]{0.65\textwidth}{
\includegraphics[width=0.65 \textwidth]{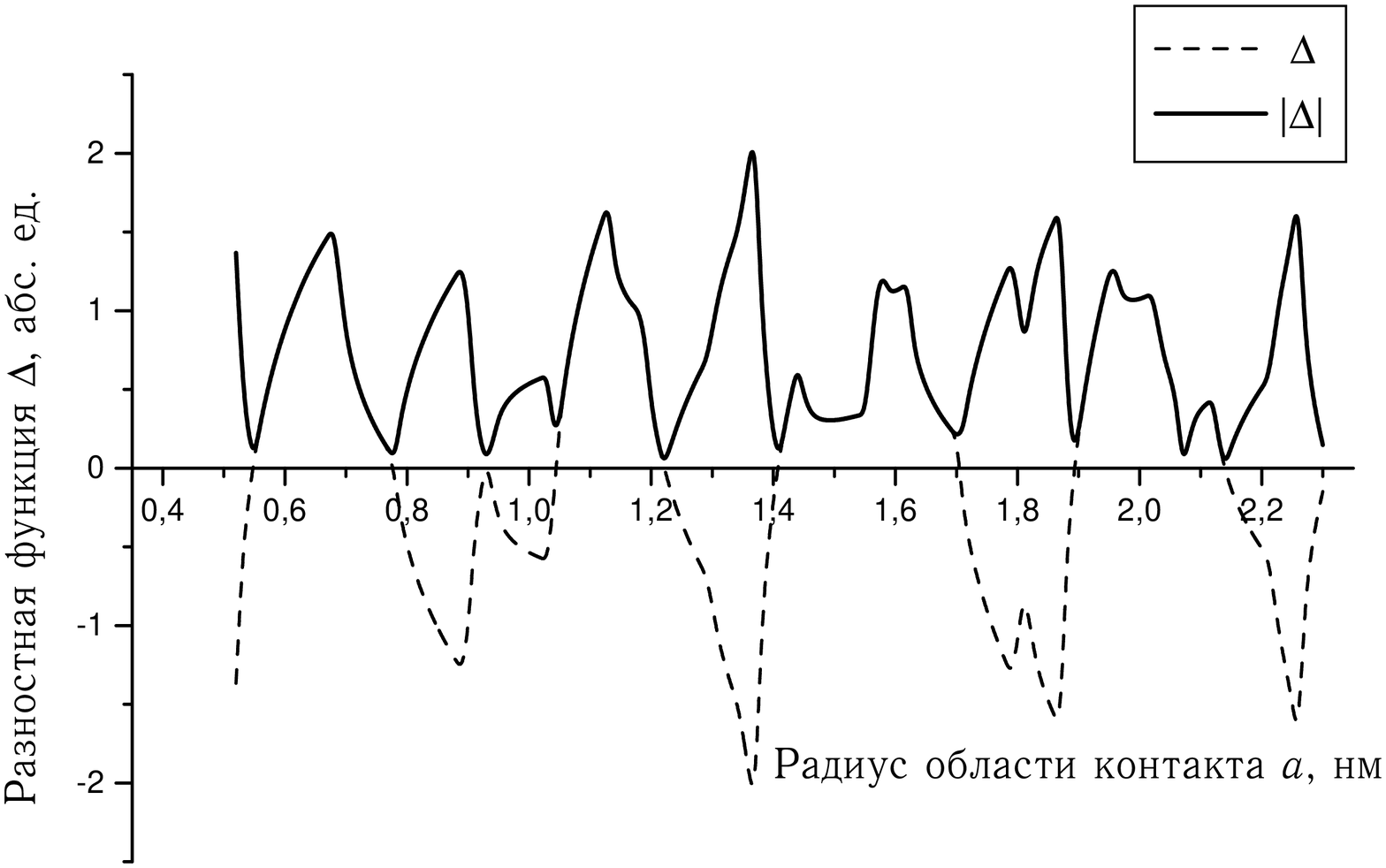}
}
\parbox[c]{0.34\textwidth}{
\caption{Зависимость \emph{разностной} функции АСМ-изо\-бра\-же\-ния~(\ref{appar6_}) и ее модуля от радиуса области контакта~$a$ для модельного случая гексагональной решетки ($d=0{,}52\,\mbox{нм}$) с $\delta$-функ\-ци\-ями в качестве атомных.}
\label{S_vid}
}
\vskip -0.75cm
\end{center}
\end{figure}

Мы провели простейший анализ, вычисляя $f$ по~(\ref{appar3}) для двух случаев: когда центр зонда находится \emph{над атомом} и \emph{между атомами} решетки. Поверхностную решетку образца описывали как гексагональную с параметром $d$, в качестве атомных функций выбирали $\delta$-функции. Используя~(\ref{appar2}) и (\ref{appar3}), обозначая для двух случаев вычисляемые функции как $f_1$ и $f_2$ и определяя \emph{разностную} функцию АСМ-изображения:
\begin{equation}
\label{appar6_}
\Delta=(f_1-f_2),
\end{equation} 
рассчитали ее зависимость от $a$ для $d=0{,}52\,\mbox{нм}$ (параметр решетки слюды), см. рис.\,\ref{S_vid}.
Как следует из рисунка, значения \emph{разностной} функции существенно варьируются в зависимости от параметра $a$. Более того, при различных значениях $a$ может изменяться знак $\Delta$, что должно приводить к наблюдению инвертированного АСМ-изображения (инверсия контраста). Особо следует подчеркнуть, что \emph{не наблюдается} тенденция к уменьшению значений максимумов разностной функции по мере увеличения области контакта и, следовательно, вовлечения в контакт все большего количества атомов; т.е. сохраняется возможность получения \lk атомного\pk\ разрешения.

\begin{figure}[!h]
\begin{center}
\includegraphics[width=0.65\textwidth]{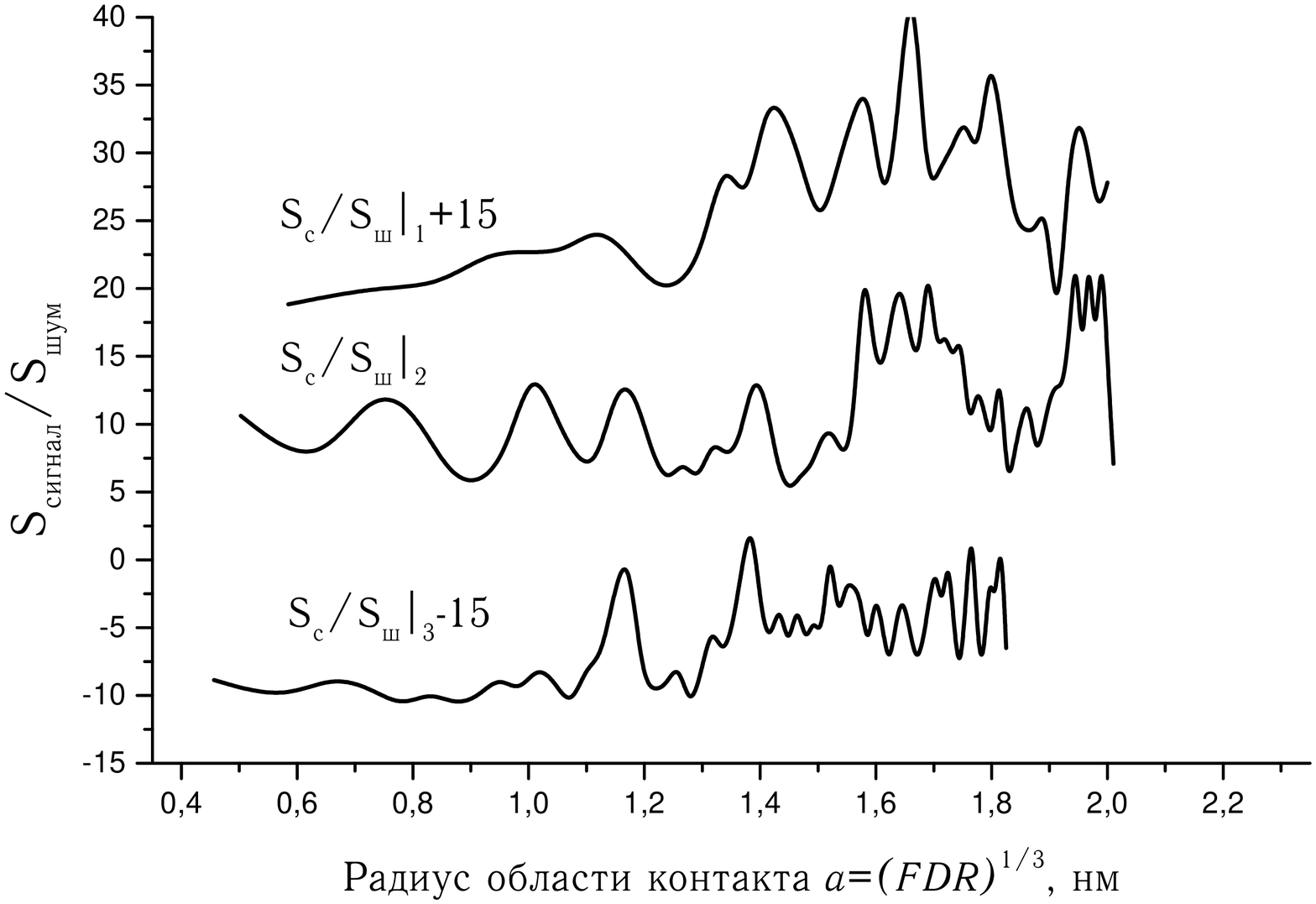}
\parbox[b]{0.34\textwidth}{
\caption{Зависимость спектральной плотности сечения функции АСМ-изо\-бра\-же\-ния слюды на пространственной частоте $k\sim 1/0{,}52\,\mbox{нм}^{-1}$ вдоль основной оси решетки от радиуса области контакта~$a$. Измеренная величина нормирована на спектральную плотность шума на той же частоте. Зависимости получены для трех различных зондов.}
\label{vid_exp1}
}
\vskip -0.5cm
\end{center}
\end{figure}

Для сравнения с экспериментом проводили измерения зависимости от радиуса области контакта спектральной плотности сечения функции $F_{тр}(x,y)$ АСМ-изображения поверхности слюды вдоль основной оси решетки на пространственной частоте $k\sim 1/0{,}52\,\mbox{нм}^{-1}$, см. графики рис.\,\ref{vid_exp1}. Измеряемая величина по своему физическому смыслу должна коррелировать с введенной выше \emph{разностной} функцией (ее модулем, рис.\,\ref{S_vid}). Эксперимент показывает, что при увеличении радиуса области контакта и увеличении числа контактирующих с зондом атомов образца действительно сохраняется возможность визуализации двумерной периодической структуры. 

Однако, согласно проведенному анализу, визуализируемая \lk атомная\pk\ структура\footnote{с теми же параметрами решетки, что и исследуемая поверхность} не является \lk истинной\pk\footnote{при типичных условиях АСМ-эксперимента на воздухе}: структура элементарной ячейки на АСМ-изображении может отображаться неадекватно (инверсия контраста); влияние точечного дефекта перераспределяется по области размера $\sim a$. Отличие предлагаемого механизма визуализации атомной структуры в АСМ от описанных в литературе состоит в том, что он позволяет связать особенности (контраст, качество) АСМ-изображения атомной структуры с реальными параметрами эксперимента (силой воздействия зонда, модулями упругости зонда и образца, радиусом кривизны иглы и степенью ее асимметрии), а не с абстрактным \lk количеством атомов на кончике иглы\pk.

\section*{Восстановление истинной геометрии объектов по АСМ-изображению (учет эффекта уширения)}
\label{width}

Эффект уширения проявляется в том, что у АСМ-изображений одиночных объектов\footnote{адсорбированных на поверхность подложки} завышены значения ширины профиля, и связан с тем, что зонд микроскопа имеет конечный радиус кривизны кончика. Традиционно для учета эффекта уширения объект исследования описывают \emph{сферической} геометрией. Мы обобщили этот подход для модели \emph{эллипсоидального} объекта (в сечении~\emdash эллипс с полуосями $a$ и $b$), что представляется существенным в свете изложенных выше результатов о контактных деформациях образцов под действием зонда.

\begin{figure}[!h]
\begin{center}
\includegraphics*[width=0.3 \textwidth]{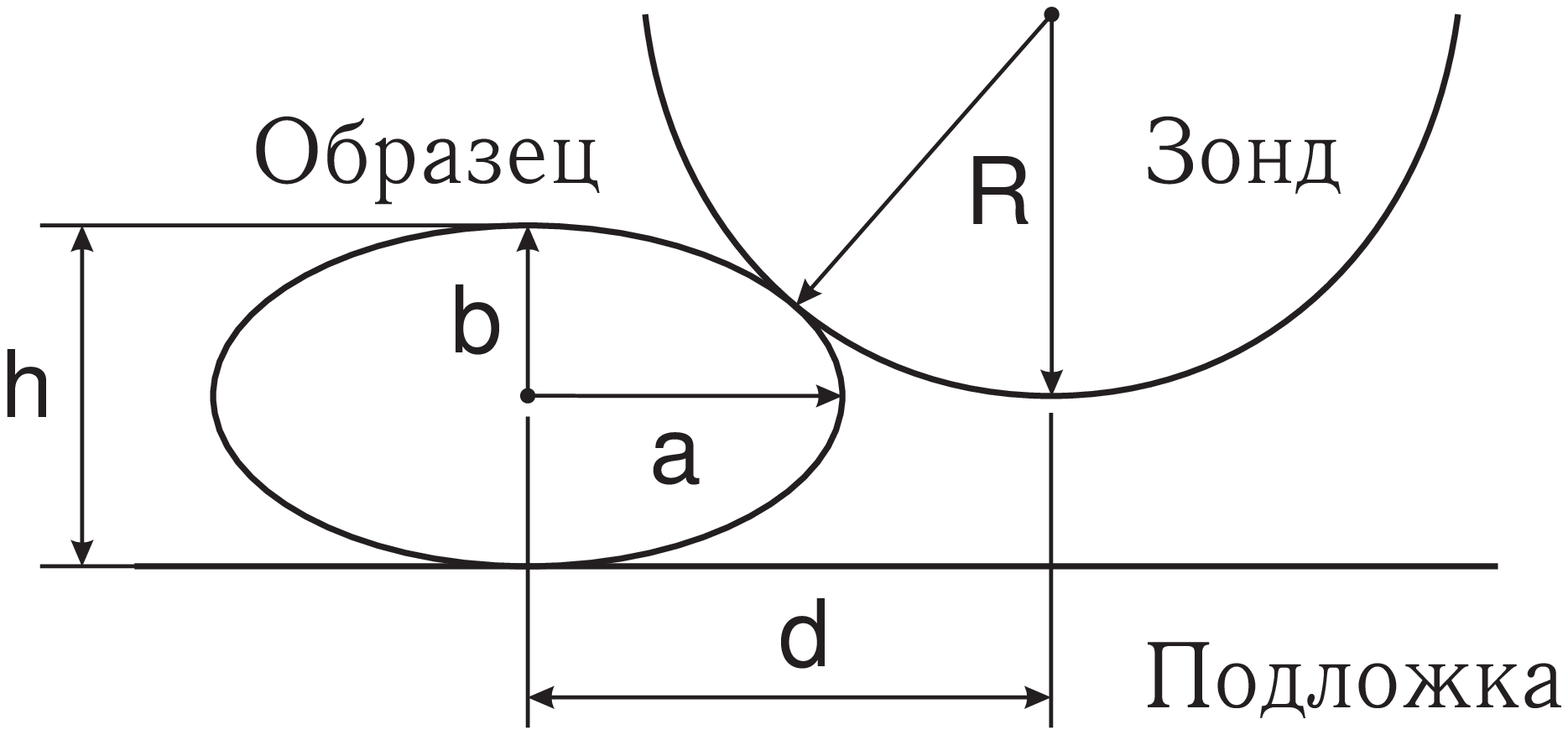}\quad
\parbox[b]{0.6\textwidth}{
\caption{Геометрия контакта зонда и образца (к объяснению эффекта уширения)}
\label{geom1}
}
\vskip -0.5cm
\end{center}
\end{figure}

Была рассмотрена геометрическая модель (рис.\,\ref{geom1}), учитывающая взаимодействие объекта только с кончиком зонда (\lk низкий\pk\ объект). Кончик зонда аппроксимировали либо полусферой радиуса $R$, либо параболоидом вращения ($z=k\rho^2$, где $k$\,\emdash коэффициент параболы), при этом было показано, что результаты применения двух методик фактически тождественны\footnote{количественная разница найденных решений составляет <10\%}. Ставили задачу: по заданным значениям высоты АСМ-профиля ($2b$), параметра геометрии иглы ($R$ или $k$) и ширины профиля АСМ-изображения на полувысоте ($2d$) определить истинное значение ширины объекта~$2a$. Анализ геометрии контакта позволил в обоих случаях свести задачу к одному уравнению с одним неизвестным, для которого был создан алгоритм численного решения. Причем показали, что полученное уравнение \emph{не имеет решения} в случае, когда выполняется условие (для модели параболической иглы):
\begin{equation}
\label{ex2}
k< b/d^2
\end{equation}
(сходное соотношение было получено и в модели сферической иглы). Смысл данных ограничений очевиден: если измеренные АСМ-профили объектов \lk острые\pk, то и игла, с помощью которой они были прописаны, также должна быть достаточно \lk острой\pk. 

Как показывает анализ устойчивости решения, определяющий вклад в погрешность $a$ вносит погрешность параметра $d$; типичные значения стандартного отклонения для найденных значений~$a$ могут превышать 20\%, поэтому для уменьшения ошибки необходим набор достаточной статистики.

\section{АСМ-исследования взаимодействия вирусной РНК с белками}
\label{chap3}

\mbox{}\footnote{Результаты, изложенные в данной главе, получены в ходе совместной работы с научной группой д.\,х.\,н. Ю.\,Ф.\,Дрыгина (НИИ ФХБ им.\,А.\,Н.\,Белозерского МГУ)} Выбор объекта исследования объясняется следующим. Вирусы являются простейшими природными объектами, способными к репликации, и состоят из носителя генетической информации (молекулы нуклеиновой кислоты) и белковой оболочки. Процесс репликации регулируется спецификой взаимодействия молекулы нуклеиновой кислоты с белками. Наблюдение свободной РНК позволяет исследовать это взаимодействие, что может дать вклад в понимание механизмов процесса сборки вирусов, структурной организации и функционирования рибонуклеопротеидов.

В настоящее время разработан ряд методик фиксации макромолекул ДНК на поверхностях подложек для АСМ-исследований. Однако в литературе отсутствовали работы, посвященные подобным экспериментам с высокомолекулярной \emph{однонитевой} РНК, что объясняется проблемой ее фиксации на подложке в расправленном состоянии. В работе апробировали методику приготовления вирусной РНК для АСМ-исследования и провели визуализацию стадий процесса последовательного высвобождения РНК из белковой оболочки частиц вируса табачной мозаики под действием химических реагентов. Следует отметить, что разработанная методика не включает какого-либо дополнительного контрастирования молекул РНК, в отличие от препарирования нуклеиновых кислот для электронной микроскопии (ЭМ).

\begin{figure}[!h]
\begin{center}
\includegraphics[width= 0.5 \textwidth]{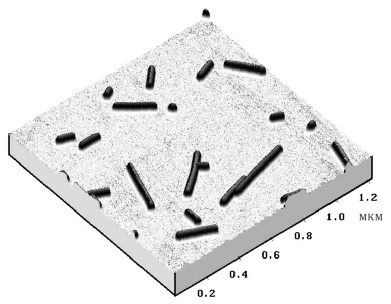}\qquad
\parbox[b]{0.44\textwidth}{
\caption{Частицы вируса табачной мозаики, адсорбированные на пирографит; АСМ-исследование}
\label{vtm_free_n}
}
\vskip -0.3cm
\end{center}
\end{figure}

Цельные частицы ВТМ были исследованы в режимах постоянного и прерывистого контактов на поверхностях пирографита и слюды (рис.~\ref{vtm_free_n}). Было обнаружено, что скорость сорбирования вирусных частиц на подложку тем выше, чем больше степень гидрофобности ее поверхности; мы связываем это наблюдение с проявлением \emph{гидрофобного} эффекта.

\begin{figure}[!h]
\begin{center}
{\footnotesize
а) \includegraphics[width= 0.35 \textwidth]{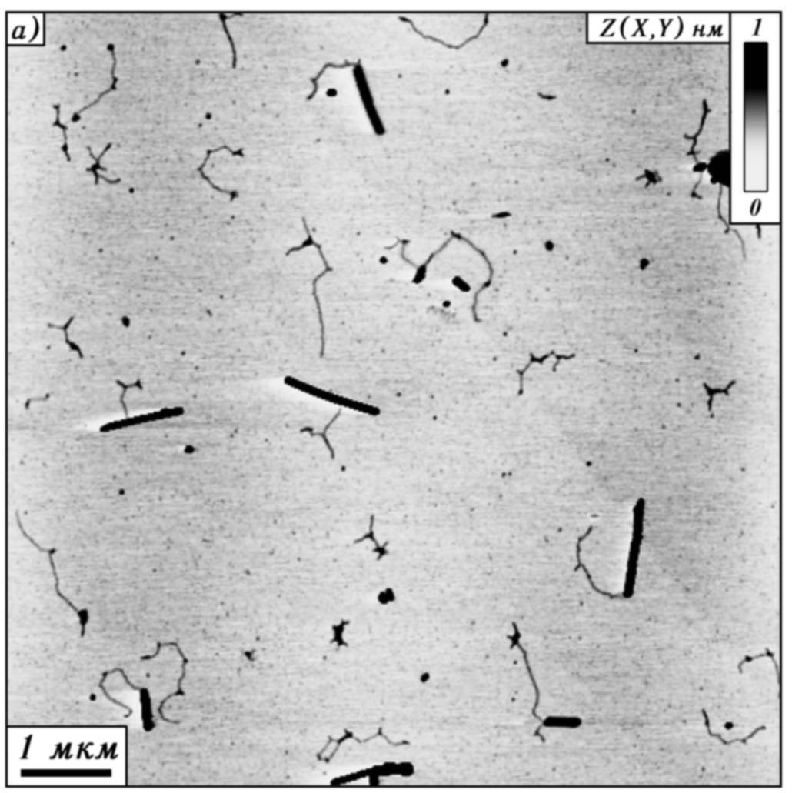}\quad
б) \includegraphics[width= 0.35 \textwidth]{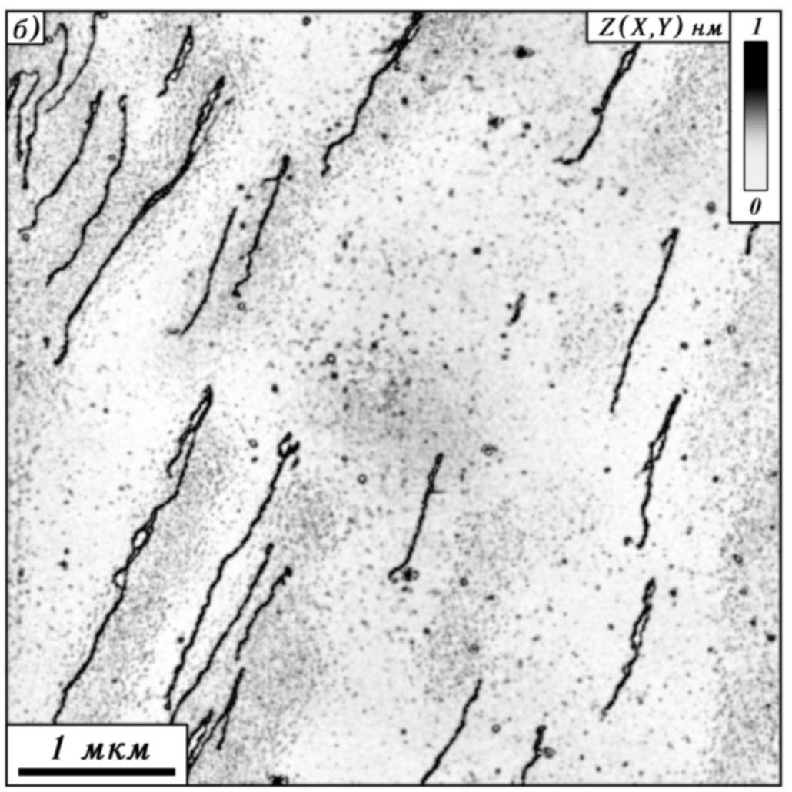}
}
\caption[Вирусные частицы мозаичной болезни табака]{а) Частично разрушенные рибонуклеопротеиды ВТМ, наблюдаются выходящие из концов частиц нити РНК; б) полностью высвобожденные молекулы РНК; АСМ-исследование, подложка~\emdash слюда}
\label{vtm_free}
\vskip -0.5cm
\end{center}
\end{figure}

	В мягких условиях депротеинизации наблюдали рибонуклеопротеиды (РНП) с выходящими из них (с одного конца или с обоих) нитями РНК, см. рис.\,\ref{vtm_free}а. Протестировав три известные методики частичной депротеинизации ВТМ и высвобождения РНК, описанные в литературе, обнаружили, что наиболее перспективным является метод с применением диметилсульфоксида (ДМСО). Метод депротеинизации в растворах со щелочным значением рН характеризуется большим фоном получаемых АСМ-изображений. Кроме того, молекулы РНК в этом случае склонны к агрегации и в значительной степени сохраняют элементы вторичной структуры. Метод с использованием мочевины позволяет получать АСМ-изображения с незначительным фоном, но при этом наблюдается агрегация вирусных РНП. 

В жестких условиях депротеинизации белковая оболочка разрушается, и макромолекула РНК полностью высвобождается (рис.\,\ref{vtm_free}б). Изображенные здесь молекулы отделены от молекул разрушенной белковой оболочки методом высокоэффективной жидкостной хроматографии и адсорбированы на свежий скол слюды из 36\% \mbox{ДМСО} в присутствии 1\% ВАС (бензилалкиламмоний хлорида). Присутствие \mbox{ДМСО} способствует расправлению молекул, а BAC~\emdash иммоблизации их на подложке.

Проводя анализ вирусных частиц, находящихся на промежуточной стадии разрушения, обнаружили, что в 40\% случаев у вирусной частицы наблюдается две выходящих из остова нити частично высвобожденной РНК (в остальных случаях~\emdash одна нить). Статистический анализ зависимости длины обеих нитей от длины вирусного остова позволил зафиксировать \emph{асимметрию} процесса высвобождения РНК из белковой оболочки для двух концов частицы, т.о. нами подтверждена полярность\footnote{впервые обнаружена по данным ЭМ} разрушения вирусных частиц табачной мозаики.

\section{Зондовая микроскопия процессов конденсации ДНК}
\label{chap4}

\mbox{}\footnote{Результаты, изложенные в данной главе, получены в ходе совместной работы с к.\,х.\,н. О.\,А.\,Пышкиной и к.\,х.\,н. В.\,Г.\,Сергеевым (Химический факультет МГУ)}
В живой природе (внутри вирусов и клеток) молекула ДНК находится в существенно \lk компактизованном\pk\ состоянии, занимая объем на несколько порядков (до $10^6$) меньший, чем в растворе\footnote{в \emph{хорошем} растворителе}. В лабораторных условиях при помощи ряда конденсирующих агентов может быть осуществлена т.н. $\psi$-компактизация\footnote{от $\psi$~\emdash \emph{psi~\emdash polymer and salt induced}} ДНК (переход клубок~$\to$ глобула); известно, что двумя \emph{основными} формами $\psi$-компактизованной молекулы являются \emph{тороидальная} и \emph{стержневая}. Применение метода АСМ для исследований компактизации молекул ДНК представляется весьма перспективным, поскольку оказывается возможным проводить исследования в нативных или близких к ним условиях, при этом можно определять не только форму, но и размеры исследуемых структур.

\section* {Определение геометрии комплексов ДНК-ПАВ, перешедших через границу раздела фаз вода/хлороформ}
\label{dna_surf}

Компактизация ДНК при взаимодействии с катионными ПАВ и прохождение комплексов ДНК-ПАВ через межфазную границу вода/малополярный органический растворитель может рассматриваться как модель транспорта генетической информации в клетку. Представлялось интересным исследовать морфологию комплексов, перешедших из воды в хлороформ, и оценить количество молекул ДНК, входящих в комплекс, восстановив (по методике учета эффекта уширения, стр.\,\pageref{width}) истинную геометрию объектов.

\begin{figure}[!h]
\begin{center}
\includegraphics[width=0.4\textwidth]{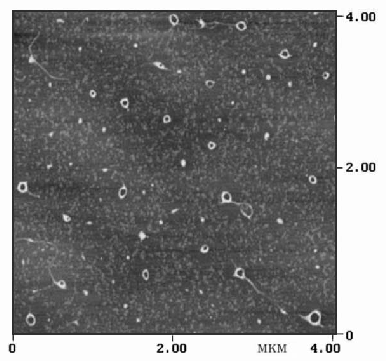}
\parbox[b]{0.5\textwidth}{
\caption[Тороидальные комплексы ДНК-ПАВ, перешедшие в хлороформ]{Тороидальные комплексы ДНК-ПАВ, перешедшие в хлороформ из водной фазы, и адсорбированные для АСМ исследования на поверхность слюды}
\label{trans_1}
}
\vskip -0.5cm
\end{center}
\end{figure}

На рис.\,\ref{trans_1} приведены результаты АСМ-визуализации комплексов ДНК-ПАВ, перешедших в хлороформ: формой комплекса является тор. С целью анализа геометрии комплексов измеряли параметры: $D$\,\emdash диаметр тора, $d$\,\emdash диаметр сечения тора, $h$\,\emdash высота тороидальной структуры над подложкой, и определили их средние значения\footnote{в качестве погрешности указаны стандартные отклонения, $N=180$}: $D=100\pm 30\,\mbox{нм}$, $d=25\pm 9\,\mbox{нм}$ и $h=5\pm 2\,\mbox{нм}$.
Для статистики пар значений $d$ и $h$ тестировали выполнение условия~(\ref{ex2}) и определяли предельное значение параметра геометрии иглы ($R$ или $k$), при котором появляются случаи отсутствия решения. Так определили, что верхняя граница значения $R$, характеризующего конкретный зонд, составляет 12 нм (т.о. путем анализа АСМ-изображений объектов получена информация и о параметрах зонда). Было рассчитано, что если радиус зонда составляет $6\div 12\,\mbox{нм}$ (нижняя граница получена из анализа тест-объектов), то среднее значение истинной ширины сечения тора $a$ составляет $10\div 11\,\mbox{нм}$ (т.о., формой комплекса ДНК-ПАВ является сплюснутый тор). Это позволило определить, что в состав 80\% комплексов входит от 2 до 16 молекул ДНК.

\section* {Исследование изменений конформации ДНК в водно-спиртовых средах}

Ранее сообщалось о результатах исследований флуоресцентной микроскопии (ФМ) динамики перехода клубок $\to$ глобула для гигантской ДНК~T4 в водно-спиртовой смеси при вариации концентрации спирта (этанол, изопропанол) в присутствии катионов Na$^+$. При этом структуру частично компактизованных образований и геометрию компактной глобулы определить не удавалось в силу ограничения разрешающей способности метода ФМ\footnote{определяется пределом разрешения оптического микроскопа}. Представлялось достаточно показательным применить метод СЗМ для исследования той же системы: визуализировать процесс в реальном масштабе времени \emph{непосредственно в водно-спиртовой среде} (в \emph{жидкостной ячейке} АСМ). 

\begin{figure}[!h]
\begin{center}
\includegraphics[width=0.75\textwidth]{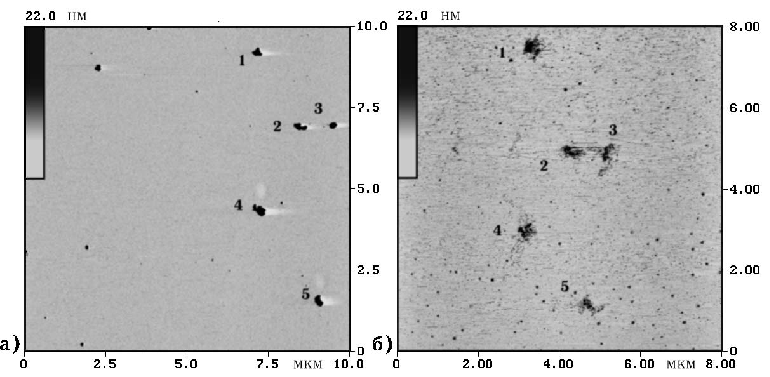} \caption[Динамика процессов компактизации/декомпактизации ДНК в водно-спиртовой среде]{Динамика процессов компактизации/декомпактизации молекул ДНК в водно-спиртовой среде; а)~\emdash компактные глобулы, выпавшие на поверхность слюды из 80\% водного раствора изопропанола: б)~\emdash частично декомпактизованные глобулы (тот же участок поверхности) после понижения концентрации спирта до 40\%; АСМ-исследование в жидкостной ячейке}
\label{dna_s_5}
\vskip -0.5cm
\end{center}
\end{figure}

Препарат, содержащий молекулы ДНК~Т4 в водно-спиртовой смеси при 80\% концентрации изопропанола, вводили в жидкостную ячейку. При этих условиях макромолекулы ДНК осаждались на поверхность слюды в виде компактных глобул (рис.\,\ref{dna_s_5}а). Применение методики учета эффекта уширения (стр.\,\pageref{width}) позволило сделать вывод, что истинной геометрической формой глобулы является сплюснутый и слегка вытянутый эллипсоид, причем каждая глобула образована одной молекулой ДНК, находящейся в достаточно плотно упакованном состоянии.

\begin{figure}[!h]
\begin{center}
\includegraphics[width=0.4\textwidth]{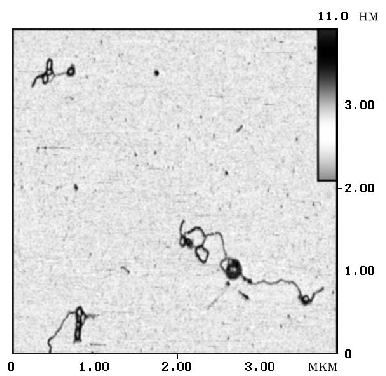} 
\parbox[b]{0.49\textwidth}{
\caption[Частично компактизованные молекулы ДНК] {Частично компактизованные молекулы ДНК, выпавшие на поверхность слюды из 40--50\% водного раствора изопропанола; АСМ-исследование в жидкостной ячейке}
\label{dna_s_6}
}
\vskip -0.5cm
\end{center}
\end{figure}

При понижении концентрации спирта в ячейке с 80\% до 40\% наблюдали частичную декомпактизацию глобулярных структур, см. рис.\,\ref{dna_s_5}б. Однако оказалось, что дальнейший процесс разворачивания глобул не происходит, что можно объяснить влиянием подложки. Поэтому макромолекулы ДНК в промежуточном состоянии (между глобулой и клубком) получали следующим образом. В жидкостную ячейку вводили молекулы ДНК, находящиеся в 40\% изопропаноле\footnote{переход клубок $\to$ глобула происходит при концентрации изопропанола более 50\%}, а затем повышали концентрацию спирта в ячейке, при этом молекулы выпадали на поверхность подложки в частично компактизованном состоянии, рис.~\ref{dna_s_6}. Было обнаружено, что начальным процессом компактизации ДНК является закручивание отдельных участков макромолекулы в тороидальные структуры, которые, по-видимому, и являются центрами дальнейшей компактизации, приводящей к формированию плотных глобул, визуализированных на рис.~\ref{dna_s_5}а.

\section{Применение метода АСМ для анализа структуры тонких органических пленок}
\label{chap5}

\mbox{}\footnote{Результаты, изложенные в данной главе, получены в ходе совместной работы с к.\,х.\,н. Г.\,К.\,Жавнерко (ИФОХ АН Беларуси, Минск)} 
Метод Ленгмюра-Блоджетт формирования органических покрытий состоит в перенесении плотноупакованных монослоев амфифильных молекул с границы фаз жидкость/газ на твердую подложку. Перенос обычно осуществляют вертикальным способом, мы использовали также метод горизонтального осаждения монослоев (при параллельности пленки и подложки). Технология ЛБ позволяет достаточно просто изменять свойства поверхности и формировать качественные пленочные покрытия с заданной структурой, что объясняет интерес к потенциальному использованию пленок ЛБ в высокотехнологичных отраслях. Однако для их широкомасштабного применения необходимо прояснить основные механизмы, определяющие структуру сформированных покрытий, и здесь новая информация может быть получена АСМ.

\begin{figure}[!h]
\begin{center}
\includegraphics[width = 0.4 \textwidth]{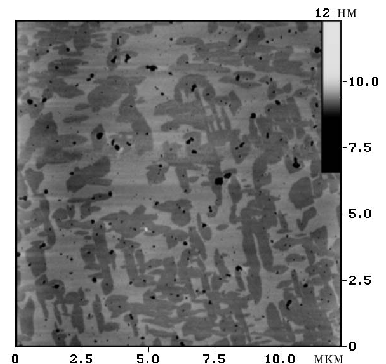}\quad
\parbox[b]{0.5\textwidth}{
\caption{Монослой бегеновой кислоты, приготовленный горизонтальным методом осаждения на слюду.\protect\\
 Визуализированы участки различного фазового состояния молекул пленки, характеризующиеся различной высотой монослоя. Наряду с этим в пленке наблюдаются поры глубиной до подложки (интенсивно темные области). АСМ-исследование}
\label{L_LS}
}
\vskip -0.5cm
\end{center}
\end{figure}

Мы обнаружили, что во многих случаях метод ГО\footnote{ГО~\emdash горизонтального осаждения} позволяет формировать более качественные покрытия, чем вертикальный метод ЛБ, что может объясняться отсутствием искажающего влияния переориентации молекул в процессе переноса в этом случае (монослой переносится на подложку \lk как есть\pk, в большей степени сохраняя свою структуру). Так, при АСМ-исследовании монослойных пленок бегеновой кислоты, сформированных на слюде методом ГО с водной субфазы, удалось визуализировать сосуществование двух областей, отличающихся высотой монослоя (разница $0{,}3\mbox{--}0{,}6$\,нм), см. рис.\,\ref{L_LS}. Мы полагаем, что области разной высоты относятся к участкам различного двумерного фазового состояния пленки (\lk твердая\pk\ и \lk жидкоконденсированная\pk\ фазы) с различной ориентацией молекул. При использовании традиционного вертикального метода ЛБ выделения монослоя визуализировать такие особенности структуры не удавалось.

\subsubsection*{Формирование кристаллитов в мономолекулярных пленках бегеновой кислоты и динамика процессов их разрушения/восстановления.}

\begin{figure}[!h]
{\footnotesize
а)\includegraphics[width = 0.4 \textwidth]{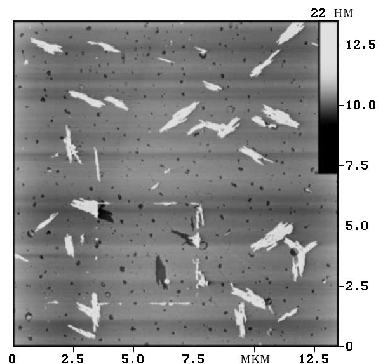}
б)\includegraphics[width = 0.4 \textwidth]{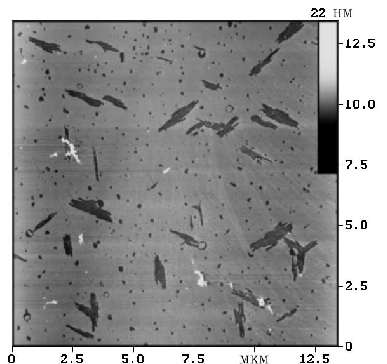}\protect\\
в)\includegraphics[width = 0.4 \textwidth]{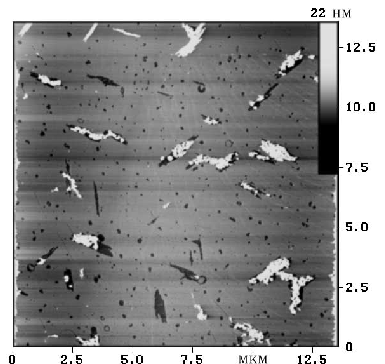}
}
\parbox[b]{0.55\textwidth}{
\caption[Кристаллиты в монослое бегеновой кислоты на слюде]{Кристаллиты, сформировавшиеся в монослое бегеновой кислоты, осажденном на подложку слюды методом ГО с субфазы, содержащей CdTe кластеры:\protect\\
 a) визуализированные при минимальном силовом воздействии зонда; б) частично разрушенные при увеличении силы воздействия зонда (более 100\,нН); в) частично восстановленные при последующей минимизации силы воздействия зонда}
\label{191crys1}
}\\

\vskip -0.5cm
\end{figure}

Мы обнаружили, что мономолекулярная пленка бегеновой кислоты, горизонтально осажденная на поверхность подложки (как слюды, так и пирографита) с субфазы, содержащей тиол-ста\-би\-ли\-зи\-ро\-ван\-ные полупроводниковые CdTe кластеры\footnote{Cd$_{54}$Te$_{32}$(SCH$_2$CH$_2$OH)$_{x \simeq 50}$~\emdash перспективный объект опто- и наноэлектроники}, содержит ламелярные кристаллические включения (рис.\,\ref{191crys1}а). Было показано, что их формирование имеет место лишь при наличии в субфазе CdTe кластеров.

Однако доказать методом АСМ присутствие кластеров в сформированном покрытии не удалось. Тем не менее, представлялось интересным исследовать структурные особенности кристаллитов. Было установлено, что ламели ориентированы не произвольным образом, а вдоль некоторых \emph{выделенных} направлений решетки подложки. Так, углы между ламелями кратны 60$^\circ$, как при использовании подложки из слюды, так и графита. Оказалось, что на подложке из слюды ламели ориентированы к основным осям решетки подложки под углом, близким 20$^\circ$, а на пирографитовой подложке~\emdash близким $30^\circ$. При этом вектор $\vec{a}$ молекулярной решетки поверхности кристаллитов (ЦП\footnote{ЦП~\emdash центрированная прямоугольная}-тип, см. ниже) ориентирован по направлению ламели.

Было обнаружено, что эти кристаллические образования разрушаются при увеличении воздействующей силы зонда (до 100~нН). После нескольких сканирований с увеличенной силой воздействия кристаллиты \lk исчезают\pk\ с АСМ-изображения (при этом окружающий их монослой остается), см. рис.\,\ref{191crys1}б. Мы считаем, что резервуаром для удаленного материала кристаллитов является боковая поверхность зонда. Если после разрушения кристаллитов продолжать сканирование, минимизировав силу воздействия зонда, то через несколько последовательных сканирований кристаллические структуры частично восстанавливаются в тех участках поверхности, где были прежде, см. рис.\,\ref{191crys1}в. При этом восстанавливается и их высота, что свидетельствует о значительной механической стабильности этих образований.

\section*{Исследование молекулярной упаковки тонких органических пленок}
\label{mol_res}

Метод АСМ представляется весьма перспективным для исследования структуры молекулярной упаковки тонких органических пленок. Основным препятствием подобных исследований является неустойчивость пленок к локальному воздействию зонда при сканировании участков малой площади, что часто приводит к локальному разрушению покрытий. Тем не менее, для ряда пленок оказалось возможным с помощью АСМ достичь молекулярного разрешения и определить параметры элементарной ячейки. Основным фактором, ограничивающим точность подобных АСМ-измерений, является \emph{температурный дрейф}, поэтому для исключения его искажающего влияния параметры молекулярной упаковки определяли путем усреднения результатов измерений значительного числа АСМ-изображений, полученных при разной ориентации держателя левера и образца и при различном направлении сканирования. Ниже приведены значения средних арифметических для измеряемых параметров, в качестве погрешности приведены величины \emph{стандартного отклонения}\footnote{ниже $N$~\emdash число измеренных АСМ-изображений}; точность калибровки прибора мы оцениваем на уровне 2\%.

\paragraph{Бегеновая кислота.} Структурная формула: С$_{21}$H$_{43}$COOH.
\vskip -0.2cm
\begin{table}[!h]

\parbox[c]{0.55\textwidth}{
{\footnotesize 
\begin{tabular} {|p{0.02\textwidth} |p{0.07\textwidth} |p{0.12\textwidth} |p{0.12\textwidth} | p{0.025\textwidth}|}
\hline
\Nom & под\-лож\-ка & $a$, нм & $b$, нм & $N$ \\
\hline
\hline
1 & слюда & $0{,}48\pm 0{,}03$ & $0{,}77\pm 0{,}06$ & 60\\
2 & графит & $0{,}50\pm 0{,}02$ & $0{,}78\pm 0{,}06$ & 70\\
3 & слюда & $0{,}47\pm 0{,}02$ & $0{,}88\pm 0{,}08$ & 70\\
4 & слюда & $0{,}53\pm 0{,}03$ & $0{,}77\pm 0{,}04$ &10\\
\hline
\end{tabular}
}
}
\parbox[c]{0.45\textwidth}{
\caption{Параметры двумерной решетки молекулярной упаковки тонких пленок бегеновой кислоты (ЦП-ячейка). \protect\\
\Nom 1 и \Nom 2~\emdash для поверхности кристаллитов (см. выше); \Nom 3~\emdash для монослоя, перенесенного вертикальным методом ЛБ на слюду, \Nom 4~\emdash для монослоя, горизонтально осажденного на слюду.
}
\label{m_res_tab3}
}
\vskip -0.3cm
\end{table}

Значения параметров молекулярной упаковки поверхности \emph{кристаллитов} (\Nom 1 и \Nom 2 таблицы \ref{m_res_tab3}), сформированных в монослоях бегеновой кислоты на субфазе, содержащей CdTe кластеры (см. выше), близки значениям $a=0{,}496$ и $b=0{,}785$\,нм, определенным Китайгородским\footnote{См., например, А.\,Н.\,Китайгородский. \emph{Молекулярные кристаллы.} \emdash М.:\,Наука, 1971\,г.} из принципа плотной упаковки углеводородных цепей для слоев молекул с R-подъячекой. Мы обнаружили, что близкими значениями характеризуется также молекулярная упаковка монослоя пчелиного воска на поверхности пирографита (инертная подложка). Можно предположить, что в этих случаях влияние подложки на молекулярную структуру минимально. В то же время для \emph{монослоев} бегеновой кислоты на \emph{слюде} (\Nom 3 и \Nom 4 таблицы~\ref{m_res_tab3}) параметры решетки отличаются от значений, определяемых принципом плотной упаковки углеводородных цепей, что может объясняться влиянием подложки. При использовании подложки \emph{пирографита} молекулярного разрешения достичь не удается. АСМ визуализирует иной тип структуры: материал пленки организуется в периодические образования (период $6{,}3\pm 0{,}4\,\mbox{нм}$) с углом взаимоориентации $60^\circ$ и углом к осям решетки подложки $30^\circ$; мы полагаем, что эти структуры являются полуцилиндрическими мицеллами.

\paragraph{Пленки кетоамидов.} Было получено молекулярное разрешение для перенесенных методом ГО на слюду монослойных пленок соединений:\protect\\

\noindent
\raisebox{10pt}{\Nom 5 N,N'-диоктадецилпропандиамида}
\includegraphics[height=0.6 cm]{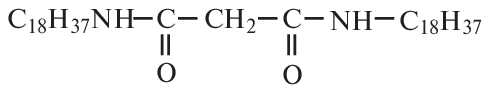}\\
\raisebox{10pt}{\Nom 6 N-гексадецилацетамида}
\includegraphics[height=0.6 cm]{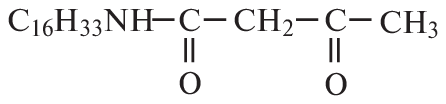}\\
\raisebox{10pt}{\Nom 7 N,N’-дигексадецилпропандиамида}
\includegraphics[height=0.6 cm]{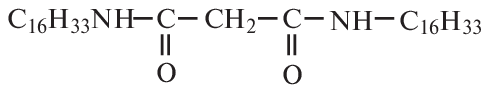}\\
\raisebox{10pt}{\Nom 8 N-гексадецил-N’-(2-нафтил)пропандиамида}
\includegraphics[height=0.6 cm]{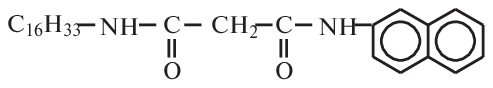}\\
\raisebox{10pt}{\Nom 9 N-гексадецил-N’-(6-хинолин)пропандиамида}
\includegraphics[height=0.6 cm]{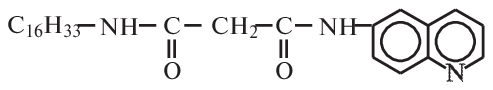}\\
\mbox{}\footnote{молекулы соединений \Nom5--7 переорганизовывались в трехслойные образования, пленки \Nom8, \Nom 9 сохраняли монослойную структуру} Было обнаружено, что за одним исключением\footnote{\Nom 8, для которого параметры решетки: $a=0{,}566\pm 0{,}015$, $b=0{,}80\pm 0{,}03$\,нм} АСМ-изображения пленок исследованных соединений (\Nom 5--7 и \Nom 9) характеризуются гексагональной решеткой с параметром ячейки, весьма близким\footnote{различие менее 1\%} значению $0{,}48$\,нм, которое характеризует упаковку слоев углеводородных цепей с H-подъячейкой (т.н. \lk газокристаллическая\pk\ фаза\footnote{другое название~\emdash \lk ротационно-кристаллическая\pk\ фаза}). Молекулы соединений \Nom 5 и \Nom 7 имеют по два углеводородных фрагмента равной длины, и можно предположить, что каждой молекуле на АСМ-изображении соответствуют \emph{две} пучности. Однако наблюдение гексагональной упаковки углеводородных цепей, попарно объединенных в молекулы, несколько неожиданно, поскольку для объяснения высокой симметрии \lk газокристаллической\pk\ фазы допускают, что углеводородные цепи сохраняют возможность независимого ротационного движения вокруг своей оси.

\paragraph{Сосуществование различных типов упаковок в структуре пленок 5-ок\-та\-де\-цил-2,4,6(1Н, 3Н, 5Н)пи\-ри\-ми\-дин\-три\-она.}

\begin{table}[!h]
\vskip -0.2 cm
\parbox[c]{0.57\textwidth}{
{\footnotesize 
\begin{tabular} {|p{0.02\textwidth} |p{0.07\textwidth} |p{0.14\textwidth} |p{0.12\textwidth} | p{0.025\textwidth}|}
\hline
\Nom, тип & под\-лож\-ка & $a$, нм & $b$, нм & $N$ \\
\hline
10 & слюда & $0{,}548\pm 0{,}009$ & $1{,}56\pm 0{,}08$ & 20\\
11 & слюда & $0{,}549\pm 0{,}009$ & $1{,}31\pm 0{,}06$ & 110\\
12 & слюда & $0{,}508\pm 0{,}007$ & $1{,}37\pm 0{,}08$ & 40\\
\hline
\end{tabular}
}
}
\parbox[c]{0.43\textwidth}{
\caption{Параметры двумерной решетки молекулярной упаковки тонких пленок 5-окта\-де\-цил-2,4,6(1Н, 3Н, 5Н)пи\-ри\-ми\-дин\-три\-она (ЦП-ячейка). \protect\\
\Nom 10~\emdash для монослоя, \Nom 11~\emdash для кристаллитов (3, 6 и 9 слоев), \Nom 12~\emdash для островковых образований (2 слоя) }
\label{m_res_tab5}
}

\vskip -0.1cm
\end{table}

\begin{figure}[!h]
{\footnotesize
а)\qquad \includegraphics[width = 0.26 \textwidth]{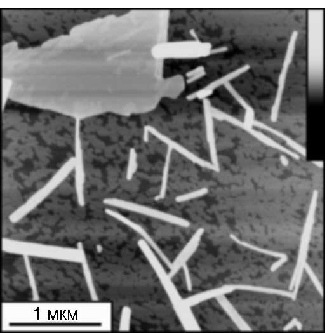}\qquad
\parbox[b]{0.54\textwidth}{
\caption[Структура тонких пленок 5-окта\-де\-цил-2,4,6(1Н, 3Н, 5Н)пи\-ри\-ми\-дин\-три\-она]{Структура тонких пленок 5-окта\-де\-цил-2,4,6(1Н, 3Н, 5Н)пи\-ри\-ми\-дин\-три\-она, АСМ-исследование. \protect\\
а)~\emdash Визуализированы области различной организации пленки: монослой, островки (2 слоя) и ламели (3, 6 и 9 слоев); б)~\emdash визуализация молекулярной упаковки монослоя и в)~\emdash поверхности кристаллитов.
}
}\\
\vskip 0.15cm
б)\qquad \includegraphics[width = 0.26 \textwidth]{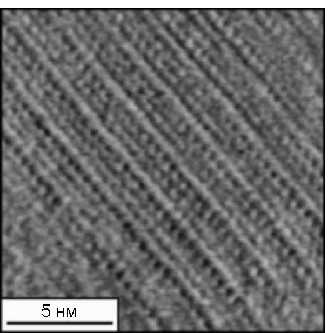}\qquad
в)\qquad \includegraphics[width = 0.26 \textwidth]{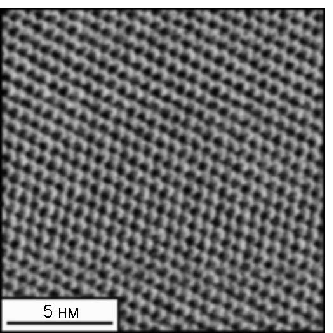} \\
}
\label{mol_res27}
\vskip -0.15cm
\end{figure}

Для этого соединения, структурная формула которого \parbox{1.75cm}{\includegraphics[height=1.2cm]{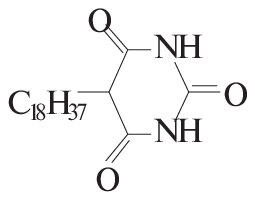}}, 
мы обнаружили три различных типа упаковки (таблица~\ref{m_res_tab5}) для участков различной структуры пленки (рис.\,\ref{mol_res27}а): монослоя, островковых образований и кристаллитов, спонтанно формирующихся в монослойном покрытии, переносимом методом ГО на слюду. 
Молекулярная упаковка островковых и кристаллических образований характеризовалась ЦП-ячейкой с отличающимися параметрами $a$ и $b$, но близкими значениями площади, приходящейся на молекулу ($0{,}35$\,нм$^2$ и $0{,}36$\,нм$^2$), см. таблицу~\ref{m_res_tab5}. В то же время АСМ-изображение молекулярной упаковки монослоя характеризовалось, во-первых, сверхструктурой (рис.\,\ref{mol_res27}б), а во-вторых, большей площадью на молекулу. Это может объясняться влиянием подложки, поскольку направление вектора $\vec b$ в этом случае совпадает с одной из кристаллографических осей слюды, причем параметр $b=1{,}56\,\mbox{нм}$ соизмерим с параметром решетки подложки~\emdash $0{,}52$\,нм (при этом параметр $a$ имеет то же значение, что и для кристаллитов). 

Т.о., мы показали, что параметры молекулярной структуры пленок определяются несколькими конкурирующими факторами: принципом плотной упаковки углеводородных цепей, значением площади полярной группы и влиянием подложки.
\clearpage

\section*{ВЫВОДЫ}

\begin{enumerate}

\item Разработаны модели и построены алгоритмы учета двух основных \emph{артефактов} АСМ: эффектов \emph{уширения} и \emph{занижения высоты} профиля одиночных объектов, адсорбированных на поверхность твердой подложки. Предложена методика определения упругих параметров отдельного микрообъекта. С позиции теории контактных деформаций проанализирован механизм АСМ-визуализации атомной структуры поверхности. 

\item Апробирована методика иммобилизации на поверхности подложки свободных молекул однонитевой вирусной РНК в расправленном состоянии. Методом АСМ визуализированы стадии процесса высвобождения РНК из белковой оболочки частиц вируса табачной мозаики и подтверждена асимметрия протекания этого процесса относительно двух концов молекулы.

\item Методом АСМ исследована динамика процесса компактизации молекулы ДНК~T4 в водно-спиртовых средах, визуализированы молекулы, находящиеся на различной стадии компактизации. Обнаружено, что частично компактизованные структуры включают тороидальные участки, образованные отдельными витками молекулы. По результатам измерений АСМ восстановлена реальная геометрия компактных структур и рассчитан их молекулярный состав.

\item По результатам АСМ-исследований структуры ЛБ пленок показана перспективность метода горизонтального осаждения монослоев на подложку. Для ряда тонкопленочных покрытий получено молекулярное разрешение и определены параметры решетки с погрешностью в единицы процентов. Показано, что структура пленки определяется совокупностью факторов: принципом плотной упаковки углеводородных цепей, значением площади полярной группы на поверхности субфазы и влиянием подложки.

\end{enumerate}

\section*{Список публикаций по теме диссертации}
\begin{enumerate}

\item V.\,Yu.\,Uvarov, Yu.\,D.\,Ivanov, A.\,N.\,Romanov, M.\,O.\,Gallyamov, O.\,I.\,Kiselyova, I.\,V.\,Yamin\-sky. Scanning tunneling microscopy study of cytochrome P-450 2B4 incorporated in proteo\-liposomes // Biochimie. \emdash 1996. \emdash V.\,78. \emdash P.\,780--784. 

\item V.\,G.\,Sergeyev, O.\,A.\,Pyshkina, M.\,O.\,Gallyamov, I.\,V.\,Yaminsky, A.\,B.\,Zezin, V.\,A.\,Kabanov. DNA-surfactant complexes in organic media // Progr Colloid Polym Sci. \emdash 1997. \emdash V.\,106. \emdash P.\,198--203.

\item М.\,О.\,Галлямов, И.\,В.\,Яминский. Нуклеиновые кислоты // Учебное пособие \lk Сканирующая зондовая микроскопия биополимеров\pk\ / под ред. д.ф.-м.н. И.В.\,Яминского. \emdash\,М.:\,Научный мир, 1997. \emdash С.\,25--40.

\item М.\,О.\,Галлямов, О.\,А.\,Пышкина, В.\,Г.\,Сергеев, И.\,В.\,Яминский. Применение методов сканирующей зондовой микроскопии к исследованию конформационных свойств ДНК // Поверхность; РСНИ. \emdash 1998. \emdash \Nom\,2. \emdash С.\,79-- 83.

\item С.\,А.\,Бычихин, М.\,О.\,Галлямов, В.\,В.\,Потемкин, А.\,В.\,Степанов, И.\,В.\,Яминский. Сканирующий туннельный микроскоп\,\emdash измерительное средство наноэлектроники // Измерительная техника. \emdash 1998. \emdash \Nom\,4. \emdash С.\,58-- 61.

\item Yu.\,F.\,Drygin, O.\,A.\,Bordunova, M.\,O.\,Gallyamov, I.\,V.\,Yaminsky. Atomic force microscopy examination of TMV and virion RNA // FEBS letters. \emdash 1998. \emdash V.\,425. \emdash P.\,217--221.

\item G.\,K.\,Zhavnerko, V.\,E.\,Agabekov, M.\,O.\,Gallyamov, I.\,V.\,Yaminsky, I.\,V.\,Lokot, F.\,A.\,Lach\-vich. AFM study of morphological peculiarities of Langmuir-Blodgett films from amphiphilic derivatives of 4-hydroxy-6-methyl-2-pyrone // Conference proceedings of third Belarussian seminar on scanning probe microscopy. \emdash\,Grodno, Oktober 1998. \emdash P.\,91--93.

\item М.\,О.\,Галлямов, О.\,А.\,Пышкина, В.\,Г.\,Сергеев, И.\,В.\,Яминский. Структура поликомплексов в водных и органических средах // Материалы всероссийского рабочего совещания \lk Зондовая микроскопия-99\pk. \emdash Н.~Новгород, март 1999. \emdash С.\,230--236. 

\item М.\,О.\,Галлямов, И.\,В.\,Яминский. Количественные методики восстановления истинных топографических свойств объектов по измеренным АСМ-изображениям: 1. Контактные деформации зонда и образца; 2. Учет эффекта уширения АСМ-профиля // Материалы всероссийского рабочего совещания \lk Зондовая микроскопия-99\pk. \emdash Н.~Новгород, март 1999. \emdash С.\,357--364; \emdash С.\,365--371

\item М.\,О.\,Галлямов, Ю.\,Ф.\,Дрыгин, И.\,В.\,Яминский. Визуализация РНК и рибонуклеопротеидов вируса табачной мозаики методами атомно-силовой микроскопии // Поверхность; РСНИ. \emdash 1999. \emdash \Nom\,7. (в печати)

\item Г.\,К.\,Жавнерко, Т.\,А.\,Кучук, В.\,Е.\,Агабеков, М.\,О.\,Галлямов, И.\,В.\,Яминский. Свойства и структура мономолекулярных пленок на основе N-окта\-де\-цил-3,4:9,10-пе\-ри\-лен\-бис(ди\-кар\-бо\-кси\-ди\-и\-мида) // Журнал физической химии. \emdash 1999. \emdash \Nom\,7. (в печати)

\item А.\,С.\,Андреева, М.\,О.\,Галлямов, О.\,А.\,Пышкина, В.\,Г.\,Сергеев, И.\,В.\,Яминский. Морфология комплексов ДНК-ПАВ, перешедших через границу раздела фаз вода-хлороформ, по результатам атомно-силовой микроскопии // Журнал физической химии. \emdash 1999. \emdash \Nom\,9. (в печати).

\end{enumerate}
см. также раздел \lk Апробация работы\pk.

\end{document}